\documentclass[reprint, superscriptaddress, twocolumn, amsmath, amssymb, aps, prb]{revtex4-2}
\usepackage{graphicx}
\usepackage{dcolumn}
\usepackage{bm}
\usepackage{placeins}
\usepackage{appendix}
\usepackage{upgreek}

\usepackage{verbatim}
\usepackage[usenames,dvipsnames]{xcolor}
 \usepackage{amsmath}
 \usepackage{amsfonts}
 \usepackage{amssymb}
 \usepackage[colorlinks=true,citecolor=blue,linkcolor=red]{hyperref}

 \usepackage{bbold}     
 \usepackage[makeroom]{cancel}  
 \usepackage{multirow}    
 \usepackage[normalem]{ulem}        
 \usepackage{array}
 \usepackage{pdfpages}
\makeatletter
 \AtBeginDocument{\let\LS@rot\@undefined}
 \makeatother

\begin{document}


\title{Gate-tunable subband degeneracy in semiconductor nanowires}

\author{Yuhao Wang}
\email{equal contribution}
\affiliation{State Key Laboratory of Low Dimensional Quantum Physics, Department of Physics, Tsinghua University, Beijing 100084, China}

\author{Wenyu Song}
\email{equal contribution}
\affiliation{State Key Laboratory of Low Dimensional Quantum Physics, Department of Physics, Tsinghua University, Beijing 100084, China}

\author{Zhan Cao}
\email{equal contribution}
\affiliation{Beijing Academy of Quantum Information Sciences, Beijing 100193, China}

\author{Zehao Yu}
\affiliation{State Key Laboratory of Low Dimensional Quantum Physics, Department of Physics, Tsinghua University, Beijing 100084, China}

\author{Shuai Yang}
\affiliation{State Key Laboratory of Low Dimensional Quantum Physics, Department of Physics, Tsinghua University, Beijing 100084, China}

\author{Zonglin Li}
\affiliation{State Key Laboratory of Low Dimensional Quantum Physics, Department of Physics, Tsinghua University, Beijing 100084, China}

\author{Yichun Gao}
\affiliation{State Key Laboratory of Low Dimensional Quantum Physics, Department of Physics, Tsinghua University, Beijing 100084, China}

\author{Ruidong Li}
\affiliation{State Key Laboratory of Low Dimensional Quantum Physics, Department of Physics, Tsinghua University, Beijing 100084, China}

\author{Fangting Chen}
\affiliation{State Key Laboratory of Low Dimensional Quantum Physics, Department of Physics, Tsinghua University, Beijing 100084, China}

\author{Zuhan Geng}
\affiliation{State Key Laboratory of Low Dimensional Quantum Physics, Department of Physics, Tsinghua University, Beijing 100084, China}

\author{Lining Yang}
\affiliation{State Key Laboratory of Low Dimensional Quantum Physics, Department of Physics, Tsinghua University, Beijing 100084, China}

\author{Jiaye Xu}
\affiliation{State Key Laboratory of Low Dimensional Quantum Physics, Department of Physics, Tsinghua University, Beijing 100084, China}

\author{Zhaoyu Wang}
\affiliation{State Key Laboratory of Low Dimensional Quantum Physics, Department of Physics, Tsinghua University, Beijing 100084, China}

\author{Shan Zhang}
\affiliation{State Key Laboratory of Low Dimensional Quantum Physics, Department of Physics, Tsinghua University, Beijing 100084, China}

\author{Xiao Feng}
\affiliation{State Key Laboratory of Low Dimensional Quantum Physics, Department of Physics, Tsinghua University, Beijing 100084, China}
\affiliation{Beijing Academy of Quantum Information Sciences, Beijing 100193, China}
\affiliation{Frontier Science Center for Quantum Information, Beijing 100084, China}
\affiliation{Hefei National Laboratory, Hefei 230088, China}

\author{Tiantian Wang}
\affiliation{Beijing Academy of Quantum Information Sciences, Beijing 100193, China}
\affiliation{Hefei National Laboratory, Hefei 230088, China}

\author{Yunyi Zang}
\affiliation{Beijing Academy of Quantum Information Sciences, Beijing 100193, China}
\affiliation{Hefei National Laboratory, Hefei 230088, China}

\author{Lin Li}
\affiliation{Beijing Academy of Quantum Information Sciences, Beijing 100193, China}

\author{Runan Shang}
\affiliation{Beijing Academy of Quantum Information Sciences, Beijing 100193, China}
\affiliation{Hefei National Laboratory, Hefei 230088, China}

\author{Qi-Kun Xue}
\affiliation{State Key Laboratory of Low Dimensional Quantum Physics, Department of Physics, Tsinghua University, Beijing 100084, China}
\affiliation{Beijing Academy of Quantum Information Sciences, Beijing 100193, China}
\affiliation{Frontier Science Center for Quantum Information, Beijing 100084, China}
\affiliation{Hefei National Laboratory, Hefei 230088, China}
\affiliation{Southern University of Science and Technology, Shenzhen 518055, China}

\author{Dong E. Liu}
\affiliation{State Key Laboratory of Low Dimensional Quantum Physics, Department of Physics, Tsinghua University, Beijing 100084, China}
\affiliation{Beijing Academy of Quantum Information Sciences, Beijing 100193, China}
\affiliation{Frontier Science Center for Quantum Information, Beijing 100084, China}
\affiliation{Hefei National Laboratory, Hefei 230088, China}

\author{Ke He}
\email{kehe@tsinghua.edu.cn}
\affiliation{State Key Laboratory of Low Dimensional Quantum Physics, Department of Physics, Tsinghua University, Beijing 100084, China}
\affiliation{Beijing Academy of Quantum Information Sciences, Beijing 100193, China}
\affiliation{Frontier Science Center for Quantum Information, Beijing 100084, China}
\affiliation{Hefei National Laboratory, Hefei 230088, China}

\author{Hao Zhang}
\email{hzquantum@mail.tsinghua.edu.cn}
\affiliation{State Key Laboratory of Low Dimensional Quantum Physics, Department of Physics, Tsinghua University, Beijing 100084, China}
\affiliation{Beijing Academy of Quantum Information Sciences, Beijing 100193, China}
\affiliation{Frontier Science Center for Quantum Information, Beijing 100084, China}


\begin{abstract}

Degeneracy and symmetry have a profound relation in quantum systems. Here, we report gate-tunable subband degeneracy in PbTe nanowires with a nearly symmetric cross-sectional shape.  The degeneracy is revealed in electron transport by the absence of a quantized plateau. Utilizing a dual gate design, we can apply an electric field to lift the degeneracy, reflected as emergence of the plateau. This degeneracy and its tunable lifting were challenging to observe in previous nanowire experiments, possibly due to disorder. Numerical simulations can qualitatively capture our observation, shedding light on device parameters for future applications.

\end{abstract}

\maketitle  

Symmetry of a confined quantum system can yield degenerate eigenstates. Breaking the symmetry lifts the degeneracy. One example is semiconductor nanowires with a symmetric cross-sectional geometry. Electrons in the wire ``feel'' a symmetric potential landscape, which may hold degenerate eigenstates. Each eigenstate corresponds to a one-dimensional electron system, a subband, with its momentum oriented along the wire axis. The conductance of each subband is quantized at $2e^2/h$ for ballistic nanowires. Therefore, conductance steps in units of $2e^2/h$ are revealed by varying subband occupation, tuned by gate voltages \cite{vanWees_1988, Wharam_1988, Kammhuber2016, Zhang2017Ballistic, InAs_Gooth, Silvano_ballistic}. Two degenerate subbands would manifest in conductance transport as the absence of a quantized plateau. Breaking the symmetry of the potential profile, e.g. by applying an electric field, can lift the degeneracy and restore the missing conductance plateau.

This subband degeneracy and its tunable lifting have been barely studied in nanowire experiments. Previous works \cite{2011_NanoLett_degenerate, 2015_degenerate} have reported a missing conductance step, but without its tunability. Moreover, the step values ($\sim$ 0.001$\times 2e^2/h$ in \cite{2015_degenerate}) significantly deviate from the quantized conductance, raising uncertainty on attributing those steps to subbands due to the non-ballistic nature of devices. In addition, the temperature in Ref. \cite{2011_NanoLett_degenerate} (77-100 K) can cause significant smoothing on small features. A bias voltage ($V$) of 10 mV was applied in Refs.\cite{2011_NanoLett_degenerate, 2015_degenerate}, and the conductance was calculated as $I/V$ ($I$ is the current). This large bias sets a low resolution for degeneracy: Subbands spacing less than 10 meV would be recognized as ``degenerate'', as a missing plateau would be expected in transport. This ``degeneracy'' is simply a biasing effect, as the missing plateau can be restored if the bias is set back to zero \cite{Kammhuber2016}.

Previous ballistic nanowires have exhibited quantized plateaus at zero magnetic field \cite{Kammhuber2016, InAs_Gooth, Silvano_ballistic}, yet without observing degeneracy even for a symmetric wire geometry. A possible reason could be residue disorder that can break the potential symmetry \cite{Jelena_degeneracy}. Additionally, the use of a single gate for tuning the electro-chemical potential could also introduce asymmetry, as the gate induces an electric field and tilts the potential profile.

In this study, we report the observation of gate-tunable subband degeneracy in PbTe nanowires with nearly symmetric geometry. A dual gate design is implemented so that the electro-chemical potential and electric field can be separately tuned  through linear combinations of gate voltages.  Absence of a quantized plateau at $2e^2/h$ or $4e^2/h$ is observed in multiple devices while other plateaus are present, indicating subband degeneracy. We further apply an electric field, and demonstrate that the missing plateau can be restored. Numerical simulations can capture these findings, suggesting a link between symmetry and degeneracy. Our observation is enabled by the significant mitigation of disorder in PbTe, see Ref. \cite{CaoZhanPbTe, Jiangyuying, Erik_PbTe_SAG, PbTe_AB, Fabrizio_PbTe, Zitong, Wenyu, Yichun, Yuhao, Vlad_PbTe, Ruidong, Wenyu_Disorder, PbTe_In} for recent progress on this material.

\begin{figure*}[htb]
\includegraphics[width=0.9\textwidth]{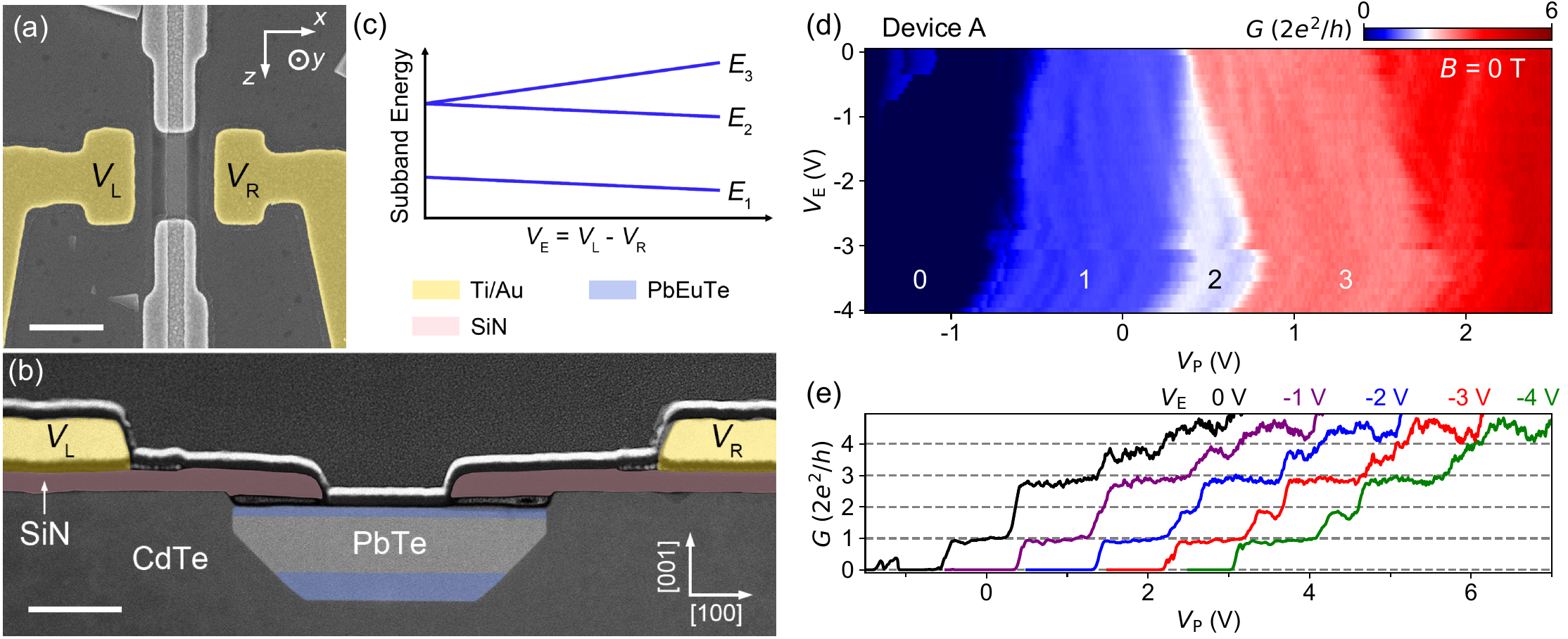}
\centering
\caption{Degeneracy and its tunable lifting in device A. (a) Device SEM. Scale bar, 500 nm. The gates are false-colored in yellow. (b) Cross-sectional image. Scale bar, 100 nm. The Pb$_{0.99}$Eu$_{0.01}$Te layers are highlighted as blue. The SiN mask is false-colored as pink. STEM of this device has been shown in our previous work for other purpose \cite{Wenyu_Disorder}. (c) Schematic of subband energy and the degeneracy lifting. (d) $G$ vs $V_{\text{E}}$ and $V_{\text{P}}$. $B$ = 0 T. (e) Line cuts from (d), with horizontal offset of 1 V between neighboring curves. }
\label{fig1}
\end{figure*}

Figure 1(a) shows the scanning electron micrograph (SEM) of a PbTe nanowire, see Ref. \cite{Wenyu_Disorder} for its growth details. The contacts and gates are evaporated Ti/Au (12 nm/43 nm). The gate voltages are denoted as $V_{\text{L}}$ and $V_{\text{R}}$, respectively. We define effective gate voltages: $V_{\text{E}}=V_{\text{L}}-V_{\text{R}}$  and $V_{\text{P}}=(V_{\text{L}}+V_{\text{R}})/2$, so that $V_{\text{E}}$ tunes the strength of the electric field, and $V_{\text{P}}$ tunes the electro-chemical potential in the wire. The cross-talk is expected to be minimal as the gate spacing's are nearly equal.

The cross-section of device A is shown in Fig. 1(b), obtained via scanning transmission electron microscopy (STEM) after its measurement. The wire has a nearly symmetric shape, and is encapsulated by CdTe substrate and Pb$_{0.99}$Eu$_{0.01}$Te (blue). The flat facets enable ballistic transport \cite{Wenyu_Disorder} and possible subband degeneracy at zero magnetic field. To facilitate the process of STEM, the whole device was covered by Ti/Au (the black and white layers on top) after the measurement.

Figure 1(d) presents the main result: Conductance, $G\equiv dI/dV$, as a function of $V_{\text{E}}$ and $V_{\text{P}}$. $V$ = 0 mV. $B$ = 0 T. The measurement circuit was two-terminal within a dilution fridge (base temperature below 50 mK). A series resistance contributed by filters and contacts (1.0 k$\Omega$) has been subtracted. At $V_{\text{E}}$ = 0 V, $G$ as a function of $V_{\text{P}}$ reveals steps near $2e^2/h$ and $3\times2e^2/h$, but lacks a distinct feature near $2\times 2e^2/h$, see the labeled numbers (in units of $2e^2/h)$ as a guidance. The black curve in Fig. 1(e) is the line cut. The absence of the ``2'' plateau suggests degeneracy of the second and third subbands ($E_2=E_3$). We denote $E_1$,  $E_2$,  $E_3$, and $E_4$ as the energies of the four lowest subbands (band bottoms).

The ``2'' plateau (white region in Fig. 1(d)) emerges and widens with an increase of $V_{\text{E}}$, indicating that the electric field lifts the $E_{2,3}$ degeneracy.  Figure 1(e) shows several line cuts with horizontal offsets. For all line cuts, see Fig. S1 in the Supplemental Material (SM). The width of the ``2'' plateau scales (roughly) linearly with $V_{\text{E}}$. A larger $V_{\text{E}}$ induces a larger electric field, leading to a stronger tilt (asymmetry) in the potential profile, yielding a larger lifting of degeneracy. Figure 1(c) is a sketch of the energy spectrum of the three lowest subbands as a function of $V_{\text{E}}$, drawn based on Fig. 1(d).

The variation of pinch-off voltages in $V_{\text{P}}$ is small across different $V_{\text{E}}$ values, suggesting that the cross-talk between them is negligibly small. $V_{\text{E}}$ mainly tunes the electric field, without substantially affecting the electro-chemical potential. The small jump after the device pinch-off in Fig. 1(e) (the black curve), also visible in the upper left corner of Fig. 1(d), is likely due to charge instabilities.

\begin{figure}[htb]
\includegraphics[width=\columnwidth]{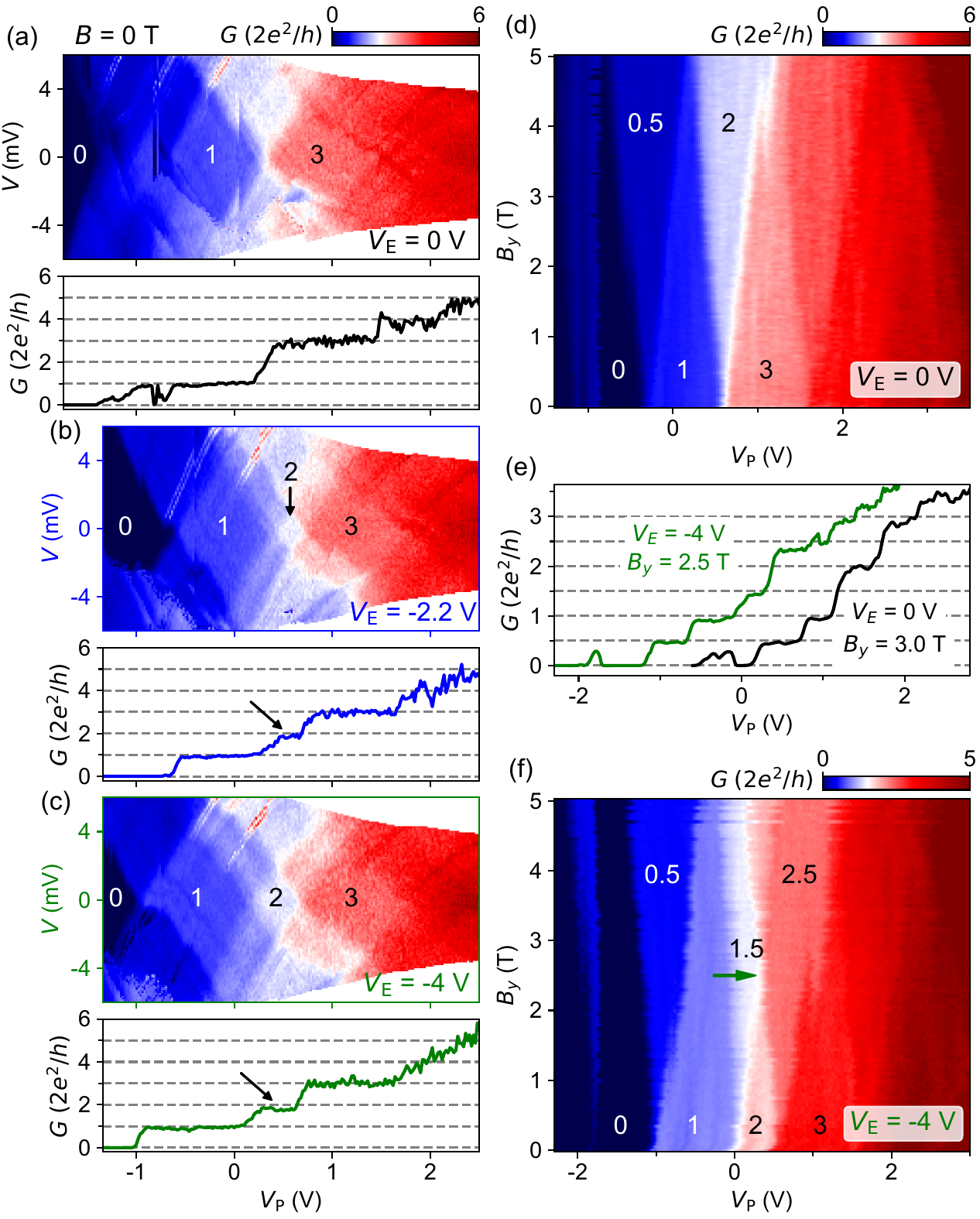}
\centering
\caption{(a-c) $G$ vs $V$ and $V_{\text{P}}$ for $V_{\text{E}}$ = 0, -2.2 and -4 V, respectively. $B$ = 0 T. Lower panels are zero-bias line cuts. All panels share the same $V_{\text{P}}$-axis. (d) $G$ vs $B_y$ and $V_{\text{P}}$ for $V_{\text{E}}$ = 0 V. $V$ = 0 V. (e) Line cuts from (d) and (f). The black curve has a horizontal offset of 0.8 V for clarity.   (f) Similar with (d) but for $V_{\text{E}}$ = -4 V. The labeled numbers indicate the plateaus (in units of $2e^2/h$). }
\label{fig2}
\end{figure}

To extract energy scales associated with subband degeneracy and its lifting, we measured the conductance map in ($V$, $V_{\text{P}}$). Figure 2(a) shows the degenerate case ($V_{\text{E}}$ = 0 V) of device A. The first and third plateaus manifest as diamond shapes in the 2D map (labeled as ``1'' and ``3''). The diamond sizes, $\sim$ 3-4 meV, correspond to $E_{2,3}-E_1$ and $E_4-E_{2,3}$, respectively. The absence of the ``2'' diamond is due to the $E_{2,3}$ degeneracy, see the lower panel in Fig. 2(a) for the line cut. The sizable dip on the ``1'' plateau is caused by a charge jump.

In Fig. 2(b), we set $V_{\text{E}}$ to -2.2 V to lift the degeneracy. The ``2'' plateau emerges as a white diamond, see the blue line cut and black arrows. The diamond size measures the amplitude of degeneracy lifting, $E_3-E_2 \sim$ 1.2 meV. Figure 2(c) further increases $V_{\text{E}}$ to -4 V, and reveals a larger diamond: $E_3-E_2 \sim$ 2.2 meV. For a rough estimation, the strength of gate-induced electric field is $(E_3-E_2)/ew \sim 7.3\times 10^3$ V/m ($w$ is averaged wire width). This strength is orders-of-magnitude smaller than the field strength induced by workfunction mismatch and accumulated surface charges \cite{CaoZhanPbTe}. $V_{\text{E}}$ thus barely modifies the direction and the strength of spin-orbit interaction, useful information for searches of helical gaps \cite{2004_PRB_Helical_theory, PRB_2014_helical_gap}.

\begin{figure}[htb]
\includegraphics[width=\columnwidth]{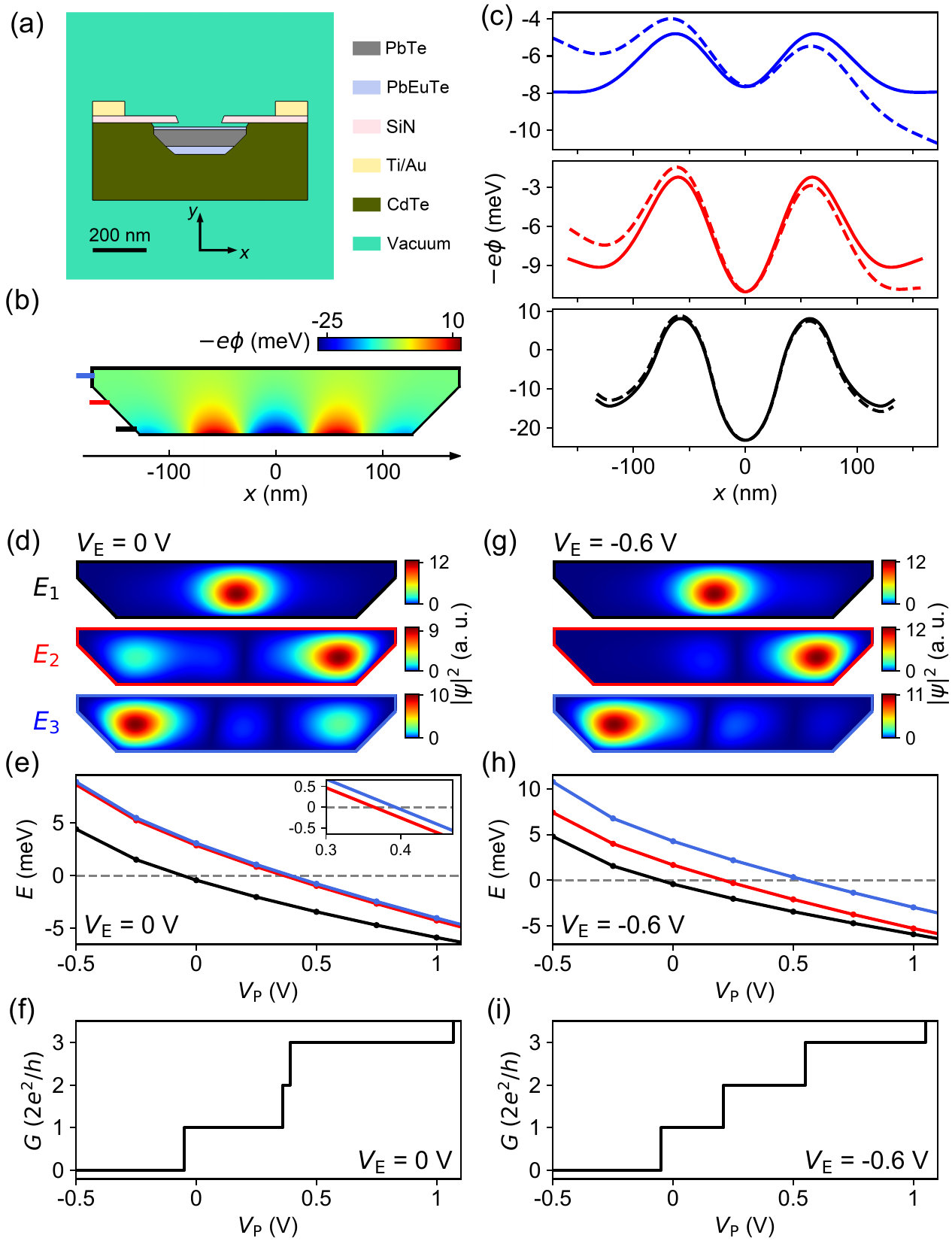}
\centering
\caption{Numerical simulations. (a) Cross-section model of device A. (b) Potential landscape in PbTe at $V_{\text{E}}$ = $V_{\text{P}}$ = 0 V. (c) Line cuts of (b) at the three color bars. The solid and dashed lines correspond to $V_{\text{E}}$ = 0 V and -0.6 V, respectively.  (d) Wavefunctions (modular square) of the three lowest eigenstates. (e) Energy of these eigenstates as a function of $V_{\text{P}}$. Inset, enlargement of the nearly degenerate point. Chemical potential $E_F$ = 0 (dashed line). (f) Plateaus in $V_{\text{P}}$ scan. $V_{\text{E}}$ = 0 V for (d-f).   (g-i) Similar to (d-f) but for $V_{\text{E}}$ = -0.6 V. }
\label{fig3}
\end{figure}

The three 2D maps in Figs. 2(a-c) quantify the energy scales related to degeneracy lifting at zero magnetic field ($B$). We next study its evolution with $B$. Figure 2(d) shows the $B_y$ scan of the degenerate case at $V_{\text{E}}$ = 0 V. $B_y$ is perpendicular to the device substrate, see the sketch in Fig. 1(a). The emergence of the ``0.5'' and ``2'' plateaus results from the Zeeman splitting of $E_1$ and $E_{2,3}$, respectively. Note that $E_{2\uparrow}$ and $E_{3\uparrow}$ remain degenerate, as do $E_{2\downarrow}$ and $E_{3\downarrow}$ ($\uparrow$ and $\downarrow$ denote the spin directions). $B_y$ does not lift the subband degeneracy, indicating that the orbital effects of $B$ on $E_2$ and $E_3$ are either identical (due to the symmetric geometry) or insignificant.  The black curve in Fig. 2(e) shows the line cut at $B_y$ = 3.0 T, resolving the ``0.5'', ``1.0'', ``2.0'', and ``3.0'' plateaus. The ``1.5'' and ``2.5'' plateaus are absent due to the aforementioned degeneracy ($E_{2\uparrow, 3\uparrow}$ and $E_{2\downarrow, 3\downarrow}$). The charge-instability-induced jumps are observable after device pinch-off.

Figure 2(f) shows the non-degenerate case at $V_{\text{E}}$ = -4.0 V, where all spin-resolved subbands can be individually revealed. The green curve in Fig. 2(e) is a line cut of Fig. 2(f) at 2.5 T. The ``2'' plateau is absent while other plateaus are present. This degeneracy arises from $B$-induced level-crossing.  As $B$ increases, $E_{2\downarrow}$ ascends and $E_{3\uparrow}$ descends in energy. They meet and cross each other at $B \sim$ 2.5 T (green arrow in Fig. 2(f)), leading to the degeneracy.  Such $B$-induced degeneracy has been observed in previous experiments \cite{Fanming_Degeneracy, DGG_Nature_Electron}, and stems from a different mechanism with the zero-field case in Fig. 2(a), despite their phenomenological similarity (the absence of ``2'' plateau). For $B$ scans along other directions, see Fig. S2 in SM.

To gain insights on the tunable degeneracy, we performed numerical simulations for device A. Figure 3(a) shows the device model, where different regions were assigned with different dielectric constants. $V_{\text{E}}$ and $V_{\text{P}}$ set the potentials of side gates (yellow), serving as boundary conditions. By solving the Poisson equation for a specific ($V_{\text{E}}$, $V_{\text{P}}$) pairing, we can obtain the potential profile, $\phi(x,y)$, inside the wire. The knowledge of surface charge accumulation, $\rho_{acc} (x,y)$, is required in this step, but cannot be obtained from experiment. In the simulations, $\rho_{acc}$ was assumed to be symmetric about $x=0$, reflecting the device's geometric symmetry. We varied its spatial distribution as a free input. With the obtained $\phi(x,y)$, we further solved the Hamiltonian to determine the eigenstates, their spacing's, and degeneracies. Simulation details can be found in SM.

Figure 3(b) shows the energy profile,  $-e\phi(x,y)$, ``seen'' by electrons in the wire at $V_{\text{E}}$ = $V_{\text{P}}$ = 0 V. This profile is obtained by assigning a non-uniform (but symmetric) $\rho_{acc}(x,y)$, see Fig. S3 in SM. The solid lines in Fig. 3(c) are horizontal line cuts of the potential profile, see the corresponding color bars in Fig. 3(b). The profile ``bends down'' near the two edges to account for accumulation of surface charges. A dip in the middle of the profile can lead to $E_{2,3}$ degeneracy, whereas $E_{1,2}$ degeneracy is associated with a peaked profile (the case of Fig. 4).  We find the non-degenerate case the most likely one by varying the profile, consistent with our observations \cite{Yuhao, Wenyu_Disorder}. Presence of dip or peak depends on the variation details in $\rho_{acc}(x,y)$. This variation can arise from the non-uniform thickness of the wire (the middle region is thicker), or an inhomogeneous environment. For example, the fabrication process may create additional surface charges in the middle region of the wire due to its exposure, but not in the side regions as they are covered by SiN mask. The dashed lines in Fig. 3(c) are the case of $V_{\text{E}}$ = -0.6 V ($V_{\text{P}}$ = 0 V). A transverse electric field, induced by $V_{\text{E}}$, tilts the profile and breaks its symmetry.

Figures 3(d) and 3(g) show the three lowest eigenstates for the profiles of solid and dashed lines in Fig. 3(c), respectively. The wavefunction of $E_1$ is mainly distributed within the middle dip, whereas $E_2$ or $E_3$ is located in the two side dips with a small coupling. Correspondingly, $E_2$ and $E_3$ are almost degenerate (Fig. 3(e)), due to the symmetric nature of the two side dips. This near degeneracy leads to the near absence of the second plateau (Fig. 3(f)). The crossings between the solid lines and the dashed line in Fig. 3(e) mark occupations of the corresponding subbands. The small splitting between $E_2$ and $E_3$ (inset of Fig. 3(e)) is due to residue overlap of their wavefunctions. The corresponding small step in Fig. 3(f) may not be observable as a plateau in transport, as the plateau visibility is also influenced by the smoothness of the saddle point potential \cite{1990_QPC_Saddle, 1994_QPC_Saddle}. For the case of $V_{\text{E}}$ = -0.6 V, the two side dips differ in energy, resulting in the lifting of $E_{2,3}$ degeneracy (Fig. 3(h)) and the emergence of the second plateau (Fig. 3(i)).

\begin{figure}[htb]
\includegraphics[width=\columnwidth]{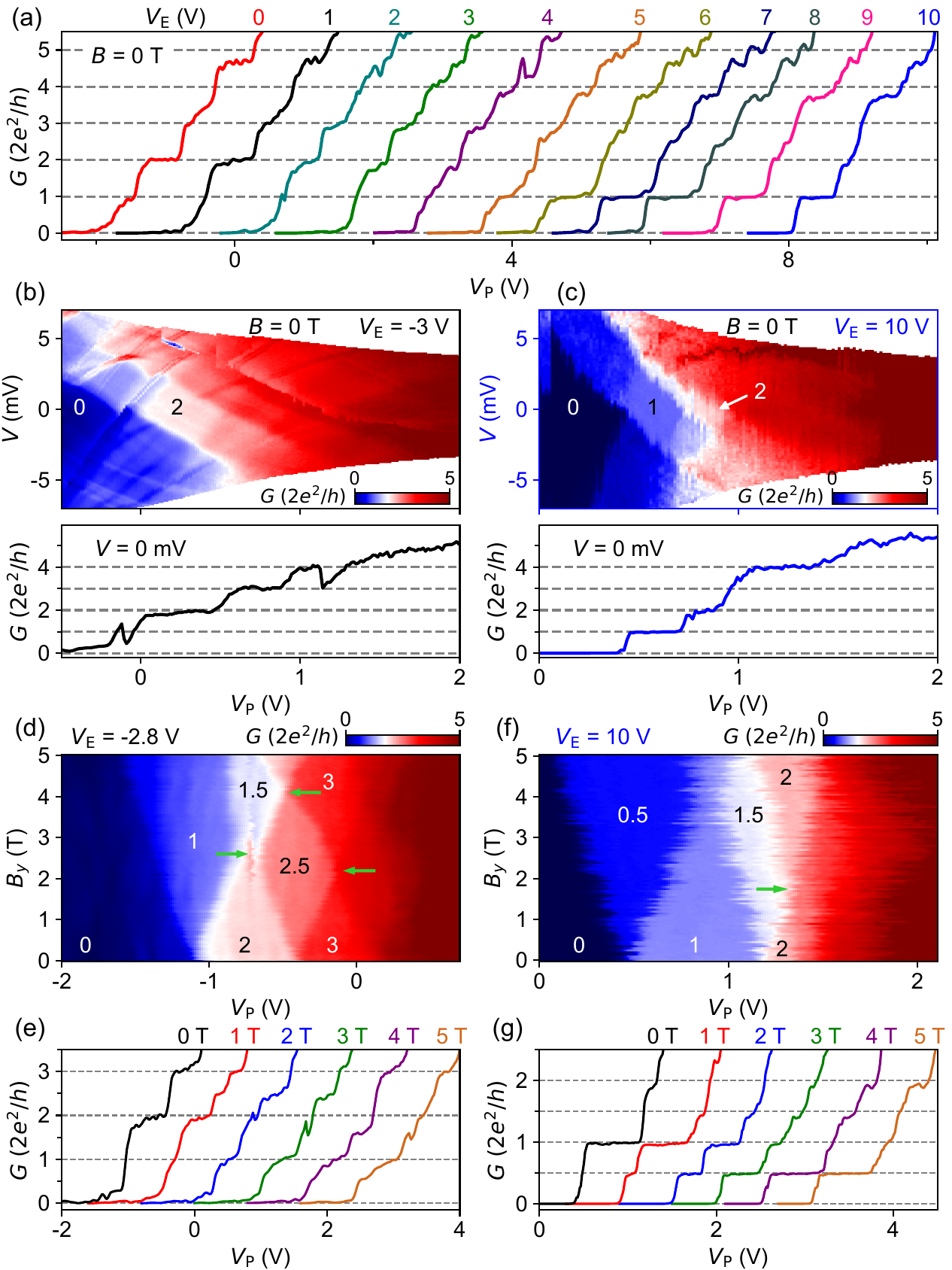}
\centering
\caption{Tunable degeneracy in device B. (a) $G$ vs $V_{\text{P}}$, with $V_{\text{E}}$ labeled. Horizontal offset between neighboring curves, 1 V. (b)  $G$ vs $V_{\text{P}}$ for $V_{\text{E}}$ =  -3 V. Lower panel, zero-bias line cut. (c) The case of $V_{\text{E}}$ =  10 V. (d) $G$ vs $B_y$ and $V_{\text{P}}$ for $V_{\text{E}}$ =  -2.8 V. $V$ = 0 V. (e) Line cuts from (d). Horizontal offset, 0.5 V. (f-g) The case of $V_{\text{E}}$ = 10 V. $V$ = 0 V.  }
\label{fig4}
\end{figure}

We next study a second device exhibiting the $E_{1,2}$ degeneracy. In Fig. 4(a), the ``1'' plateau ($2e^2/h$) is absent or barely visible at low $V_{\text{E}}$'s (0-3 V), and is restored at high $V_{\text{E}}$'s (the lifting of $E_{1,2}$ degeneracy). The gate spacing of this device is larger than that of device A, see Fig. S4 in SM for its image, thus necessitating higher $V_{\text{E}}$'s to lift the degeneracy. The slight asymmetry of wire geometry and gate spacing can be compensated by adjusting $V_{\text{E}}$ near 0 V. The (nearly) degenerate case refers to low $V_{\text{E}}$'s ($<$ 3 V). Figure 4(b) shows the 2D map of this case at $V_{\text{E}}$ = -3 V. The ``1'' diamond is absent, while the ``2'' diamond is present, suggesting $E_3-E_{1,2} \sim$ 2.1 meV. Figure 4(c) is the non-degenerate case ($V_{\text{E}}$ = 10 V). The appearance of ``1'' diamond measures the magnitude of degenearcy lifting, $E_2-E_1 \sim$ 2.6 meV. For numerical simulations of device B, see Fig. S5 in SM.

Figures 4(d-g) are the $B_y$ scans of the two cases. $B$ does not lift the $E_{1,2}$ degeneracy, as shown by the absence of the ``0.5'' plateau in Fig. 4(d). In the non-degenerate case (Fig. 4(f)), all integer and half plateaus can be revealed.  $B$-induced degeneracy (level-crossing) is present in both cases, see the green arrows. Notably, the left arrow in Fig. 4(d) marks the location where $E_{3\uparrow}$ = $E_{1\downarrow, 2\downarrow}$, signifying simultaneous $B$-induced and symmetry-induced degeneracies. The same also applies to the upper arrow, where the conductance plateau jumps by $1.5 \times 2e^2/h$. For additional data of device B, see Fig. S4 in SM. In Fig. S6, we show tunable subband degeneracy in a third device.

In summary, we have observed nearly degenerate subbands in PbTe nanowires that possess a nearly symmetric geometry. The degeneracy can be lifted by a gate-induced electric field. Numerical simulations qualitatively capture the results, suggesting a link between symmetry and degeneracy. PbTe nanowires have attracted much interest for the realization of Majorana zero modes \cite{Lutchyn2010, Oreg2010, WangZhaoyu, MS_2023}. Given that disorder has been the major roadblock in  Majorana searches \cite{Patrick_Lee_disorder_2012, Prada2012, Loss2013ZBP, Liu2017, Loss2018ABS, GoodBadUgly, DasSarma_estimate, DasSarma2021Disorder, Tudor2021Disorder}, our results, enabled by disorder mitigation, can serve as a benchmark experiment toward low-disordered Majorana nanowires. 

\textbf{Acknowledgment} This work is supported by National Natural Science Foundation of China (92065206, 12374158, 12074039) and the Innovation Program for Quantum Science and Technology (2021ZD0302400). Raw data and processing codes within this paper are available at https://doi.org/10.5281/zenodo.10912321

\bibliography{mybibfile}

\newpage

\onecolumngrid

\newpage


\section*{Supplementary material}

# Supplemental Material for "Gate-tunable subband degeneracy in semiconductor nanowires"

Yuhao Wang,[1, *] Wenyu Song,[1, *] Zhan Cao,[2, *] Zehao Yu,[1] Shuai Yang,[1] Zonglin Li,[1] Yichun Gao,[1] Ruidong Li,[1] Fangting Chen,[1] Zuhan Geng,[1] Lining Yang,[1] Jiaye Xu,[1] Zhaoyu Wang,[1] Shan Zhang,[1] Xiao Feng,[1, 2, 3, 4] Tiantian Wang,[2, 4] Yunyi Zang,[2, 4] Lin Li,[2] Runan Shang,[2, 4] Qi-Kun Xue,[1, 2, 3, 4, 5] Dong E. Liu,[1, 2, 3, 4] Ke He,[1, 2, 3, 4, †] and Hao Zhang[1, 2, 3, ‡]

[1]State Key Laboratory of Low Dimensional Quantum Physics,
Department of Physics, Tsinghua University, Beijing 100084, China
[2]Beijing Academy of Quantum Information Sciences, Beijing 100193, China
[3]Frontier Science Center for Quantum Information, Beijing 100084, China
[4]Hefei National Laboratory, Hefei 230088, China
[5]Southern University of Science and Technology, Shenzhen 518055, China

## Details of numerical simulations

We simplified the device as an infinitely long 1D nanowire (along the $z$-direction), and focused on the 2D cross-section (in the $xy$-plane). In the absence of a magnetic field, the wire Hamiltonian is $H = \frac{p_x^2}{2m_{xx}} + \frac{p_y^2}{2m_{yy}} + \frac{p_x p_y}{m_{xy}} - E_F - e\phi(x, y)$, where $p_x$ and $p_y$ are the momentum operators, $m_{xx}$, $m_{yy}$, and $m_{xy}$ are the anisotropic effective masses, $E_F$ is the Fermi energy, $e$ is the modulus of the electron charge, and $\phi(x, y)$ is the electrostatic potential tuned by gates. We neglected the Rashba spin-orbit interaction, as the gate-induced electric field inside the wire is relatively weak.

The energies of subband bottoms can be obtained by solving this 2D Hamiltonian. $\phi(x, y)$ is determined by the 2D Poisson equation: $\nabla \cdot (\epsilon_0 \epsilon_r(x, y) \nabla \phi(x, y)) = -[\rho_e(x, y) + \rho_h(x, y) + \rho_{acc}(x, y)]$, where $\epsilon_0$ is the vacuum dielectric constant, $\epsilon_r(x, y)$ is the relative dielectric constant, $\rho_e(x, y)$ is the electron density, $\rho_h(x, y)$ is the hole density, and $\rho_{acc}(x, y)$ is the accumulated surface charge density.

We used Thomas-Fermi approximation to determine $\rho_e(x, y)$ and $\rho_h(x, y)$ within the PbTe nanowire: $\rho_e(x, y) = \frac{-e\{2m_d^e[-E_c(x,y)]\Theta[-E_c(x,y)]\}^{3/2}}{3\pi^2\hbar^3}$, and $\rho_h(x, y) = \frac{e\{2m_d^h[E_v(x,y)]\Theta[E_v(x,y)]\}^{3/2}}{3\pi^2\hbar^3}$. $m_d^e = \sqrt[3]{m_l^e(m_t^e)^2}$ and $m_d^h = \sqrt[3]{m_l^h(m_t^h)^2}$ are the density-of-states effective masses (for electron and holes) with a constant energy (an ellipsoidal-shaped surface). We set $m_l^e = 0.24m_e$, $m_t^e = 0.024m_e$, $m_l^h = 0.31m_e$, and $m_t^h = 0.022m_e$ ($l$ represents the longitudinal and $t$ represents the transverse directions). $E_c(x, y) = -e\phi(x, y) - E_F$ and $E_v(x, y) = E_c(x, y) - E_g$ ($E_g = 0.19$ eV, is the band gap of PbTe bulk) are the edges of conduction and valence bands. $\Theta$ is the Heaviside step function, corresponding to the Fermi-Dirac distribution at zero temperature.

We took $\rho_{acc}(x, y)$, which is unknown in experiments, as a free input in the simulation. $\rho_{acc}$ was assumed to be symmetric about $x = 0$ for device A due to its symmetric geometry, but inhomogeneous over $x$ due to reasons mentioned in the main text.

Combining $\rho_{acc}(x, y)$, $\rho_e(x, y)$ and $\rho_h(x, y)$ with boundary conditions (set by the gate voltages), we obtained $\phi(x, y)$ by numerically solving the Poisson equation. A finite element method was used by enclosing the device within a large enough 2D rectangular box (see Fig. 3(a)). Neumann boundary conditions were imposed on the four edges of the box, while Dirichlet boundary conditions, $\phi = V_L$ and $\phi = V_R$, were imposed on the left and right gates.

We used the following parameters: $E_F = 0$, $\epsilon_r^{SiN} = 7.5$, $\epsilon_r^{CdTe} = 10.3$, $\epsilon_r^{PbTe} = \epsilon_r^{PbEuTe} = 1000$. The valley degeneracy is not observed in experiment, possibly due to the strain induced by substrate. We thus neglected valley degeneracy in the model. The effective mass, estimated based on the plateau sizes, should be smaller than the bulk values in literature. We thus used $m_{xx} = m_{yy} = 0.013$ $m_e$ (0.02 $m_e$) for device A (B), $m_{xy} = 0.2$ $m_e$, leading to subband spacing qualitatively consistent with the plateau size.

_____
* equal contribution
† kehe@tsinghua.edu.cn
‡ hzquantum@mail.tsinghua.edu.cn

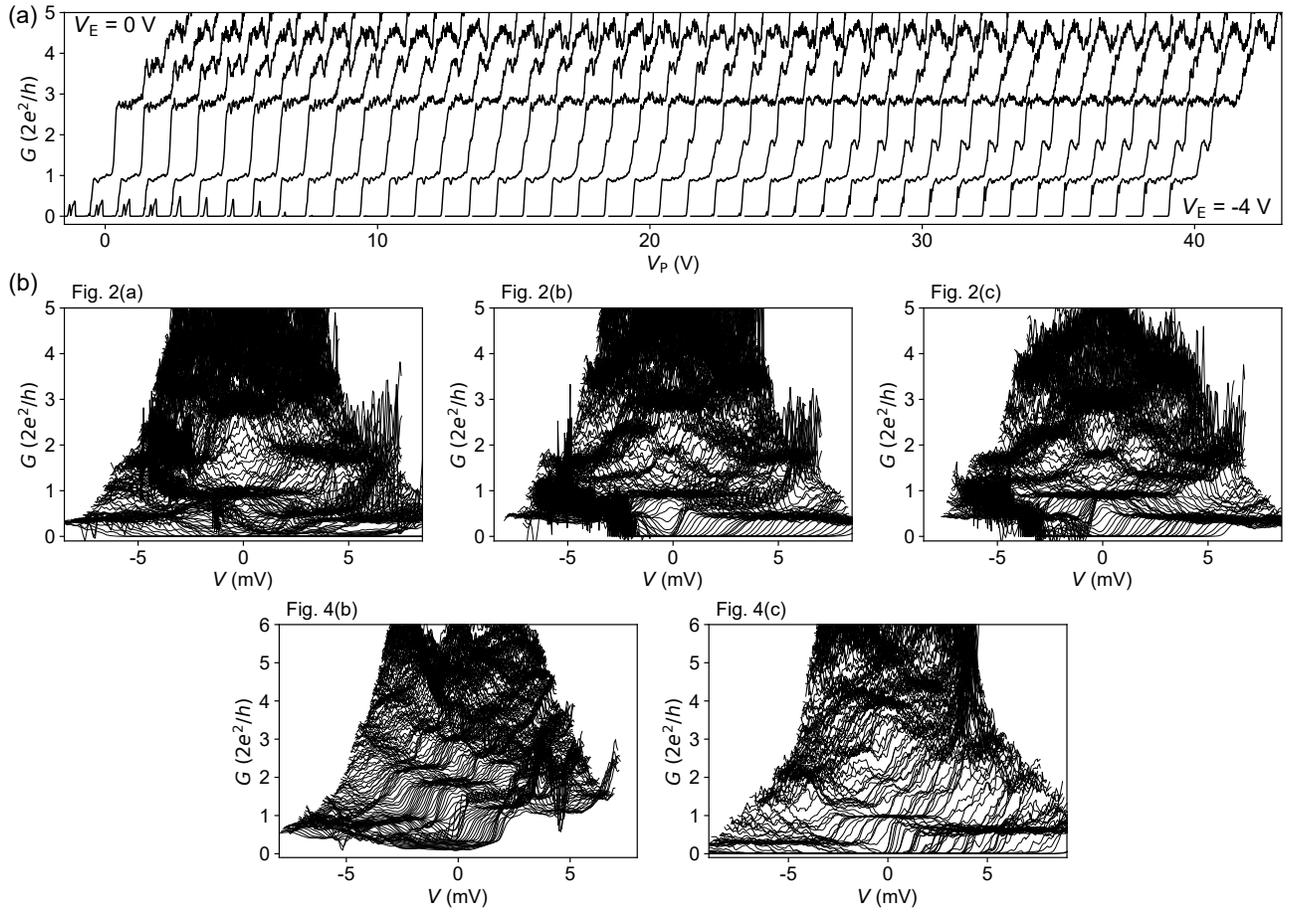

FIG. S1. (a) All line cuts of Fig. 1(d) with a horizontal offset of 1 V between neighboring curves. (b) Waterfall plots of the 2D maps in main text figures (see labeling). The plateaus are resolved as clusters of line cuts.

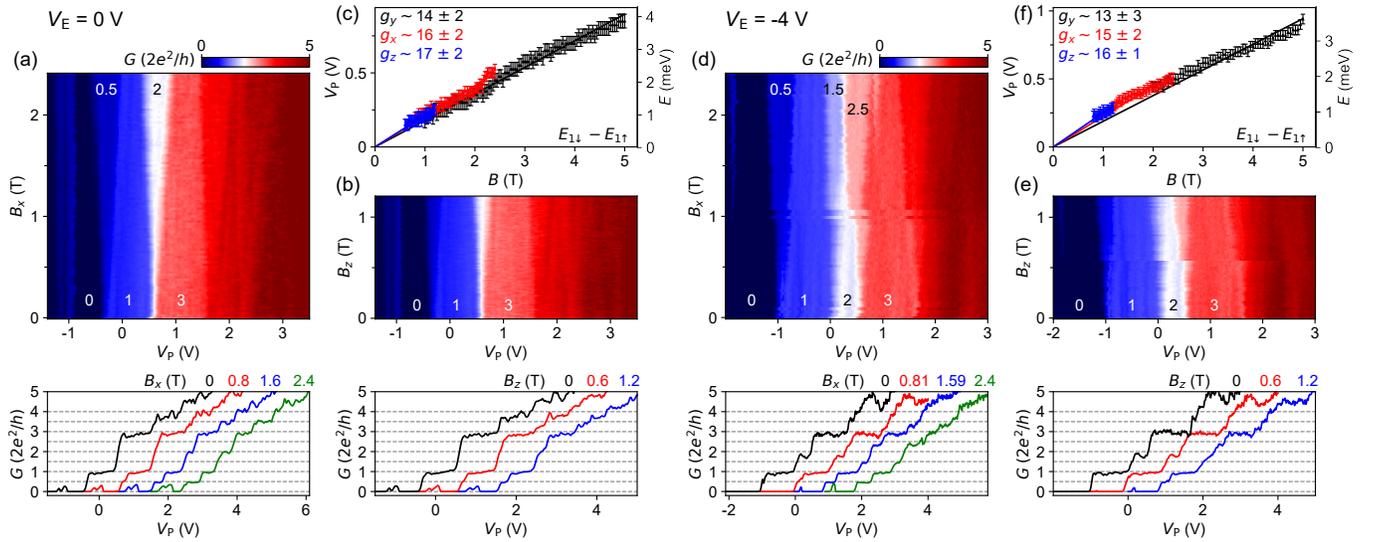

FIG. S2. $B$ scans along other directions for device A. (a) $B_x$ scan. (b) $B_z$ scan. Lower panels, line cuts (horizontal offset, 1 V). (c) $g$-factor fitting for the first subband along the three directions. To fit the width of the "0.5" plateau using the formula $g\mu_B B$, $V_P$ is converted to energy based on a lever arm extracted from Fig. 2. (a-c) are the degenerate case, $V_E = 0$ V. (d-f) Similar to (a-c) but for the non-degenerate case, $V_E = -4$ V.

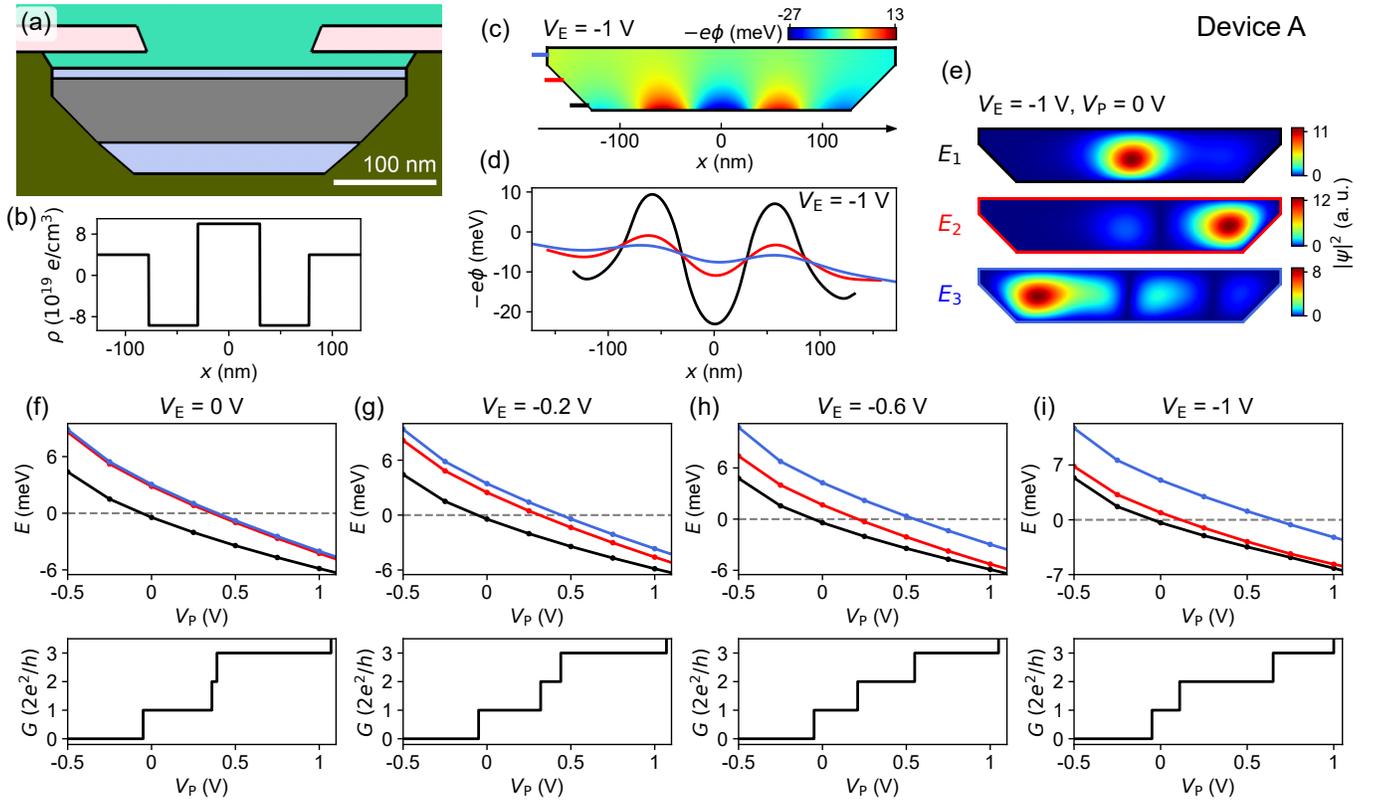

FIG. S3. Additional information on simulations of device A. (a) Device model. (b) Assigned $\rho_{acc}$. We assumed this charge distribution on the bottom surface of PbTe nanowire (within a 1-nm-thick layer). The potential profiles in Fig. 3 are generated by this $\rho_{acc}$, yielding results qualitatively consistent with experiments. Note that the actual distribution may differ from this assumption. (c) Potential profile at $V_E = -1$ V, $V_P = 0$ V. (d) Three line cuts from (c), see the colored bars. (e) Wavefunctions of the three lowest eigenstates, corresponding to the potential profile of (c). Color represents the modular square of wavefunction. (f-i) Four cases of $V_E$ illustrate the gradual evolution of the degeneracy lifting. The cases of 0 V and -0.6 V have been shown in Fig. 3. The "2" plateau becomes larger for larger $|V_E|$.

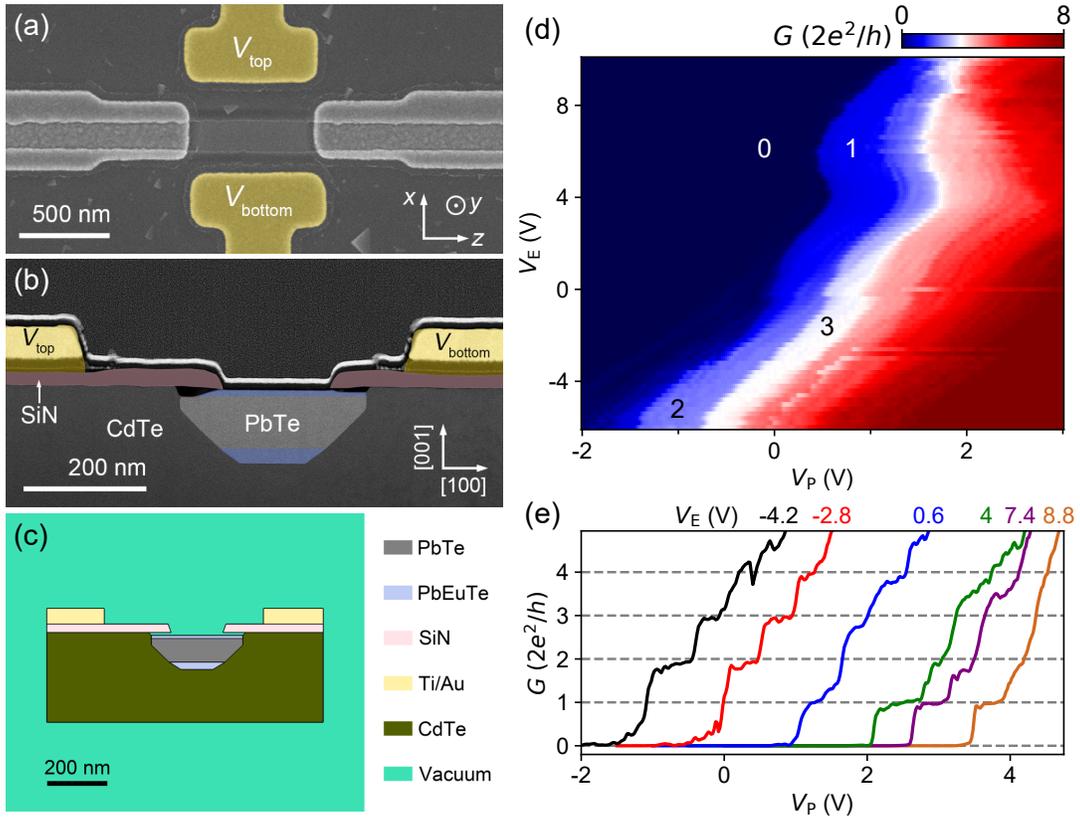

FIG. S4. Additional data of device B. (a) False-colored SEM. The bottom gate is closer to the wire compared to the top gate. Thus, a non-zero $V_E$ is needed to compensate this slight asymmetry. $V_E = V_{top} - V_{bottom}$, $V_P = \frac{1}{2}(V_{top} + V_{bottom})$. (b) Cross-sectional STEM. (c) Theory model. (d) $G$ vs $V_P$ and $V_E$ at $B = 0$ T. The labeled numbers are the plateau values in units of $2e^2/h$. For $V_E$ near -2 V, the "1" plateau is absent, suggesting the $E_{1,2}$ degeneracy. For larger $V_E$'s, the appearance of "1" plateau suggests degeneracy lifting. The pinch-off voltage in $V_P$ varies for different $V_E$'s, indicating their sizable cross-talks, possibly due to the asymmetric gate spacing's. (e) Several line cuts from (d) with a horizontal offset of 0.5 V. The line cuts in Fig. 4(a) were measured separately from (d-e). The subtracted contact resistance, $R_{contact}$, for device B is 850 Ω, except for Fig. 4(b) and Fig. 4(c) where $R_{contact}$ is 600 Ω and 750 Ω, respectively. This variation might be related to the device instability or a gate effect.

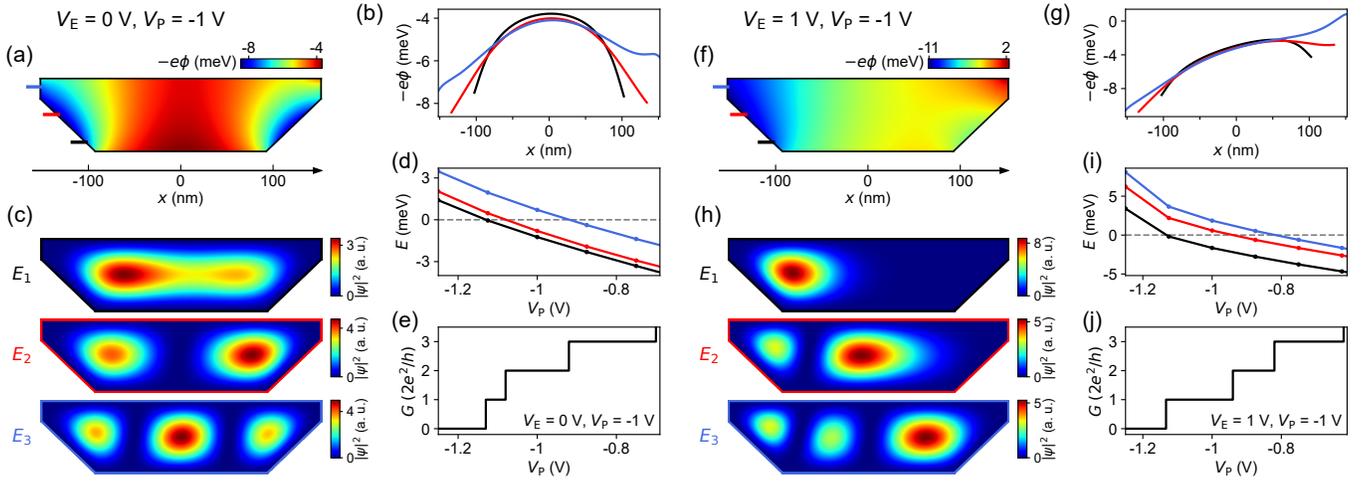

FIG. S5. Numerical simulations of device B. $V_E = 0$ V for (a-e). (a) Potential energy profile. We assumed a uniform surface charge accumulation on the left side facet, $\rho_{acc,L} = 0.7 \times 10^{19} e/\text{cm}^3$ within a 1-nm-thick layer, while for the right side facet it is $\rho_{acc,R} = 0.83 \times 10^{19} e/\text{cm}^3$. (b) Line cuts from (a), see the corresponding color bars. (c) Wavefunction (modular square) of the three lowest eigenstates. (d) Energy of $E_1$, $E_2$, and $E_3$ as a function of $V_P$. $E_1$ and $E_2$ are nearly degenerate. (e) Plateau plot. The "1" plateau is significantly smaller than the "2" plateau, and may be barely visibly in transport. Note that this plot is not transport calculation which would require a 3D model, including the source/drain contacts and a saddle-like potential profile. The visibility of plateau also depends on the smoothness of the saddle potential. (f-j) The case of $V_E = 1$ V, where the degeneracy is lifted.

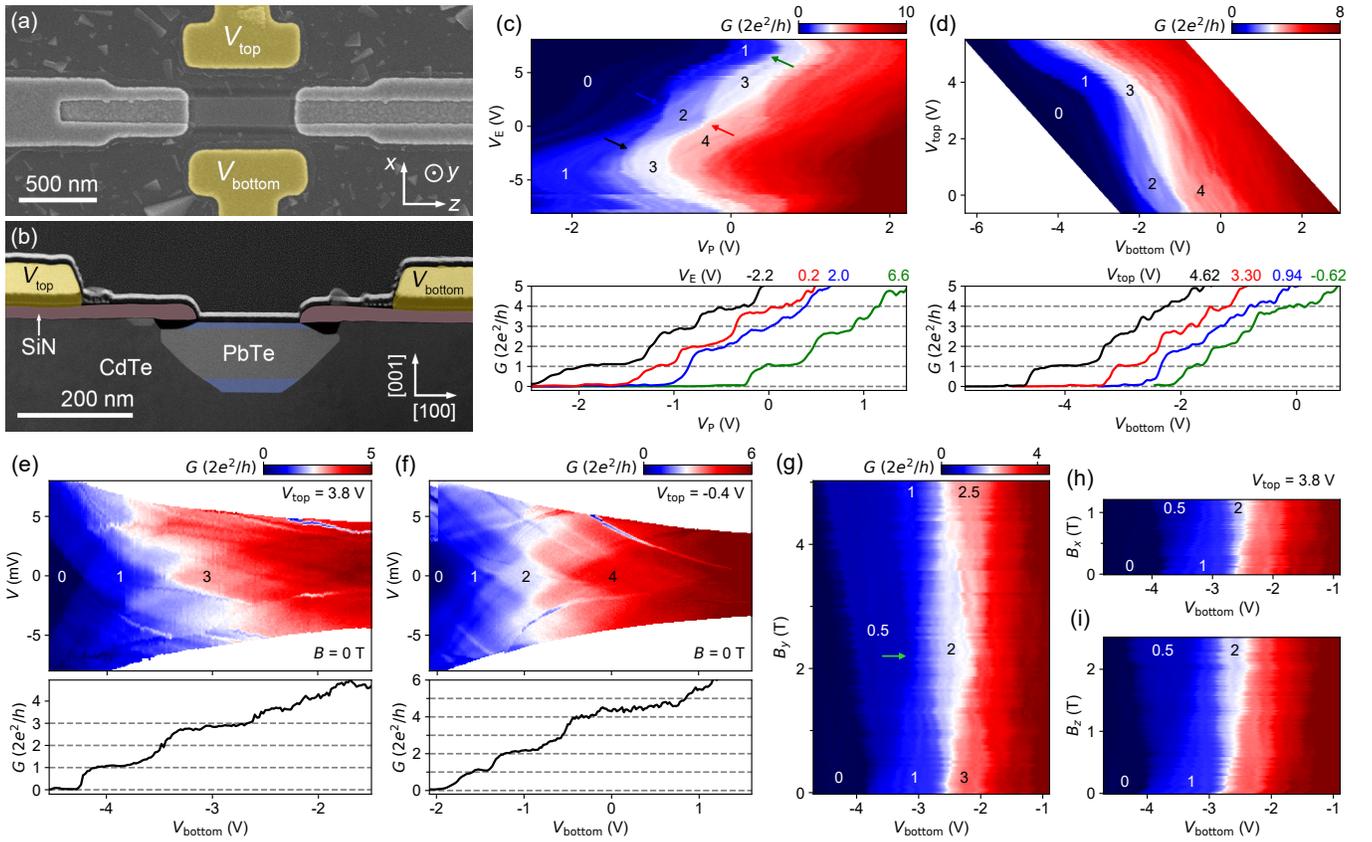

FIG. S6. Subband degeneracy in device C. (a) Device SEM. (b) Cross-sectional STEM of device C, performed after its measurement. (c) $G$ vs $V_E$ and $V_P$ at $B = 0$ T. $V = 0$ mV. Lower panel, four horizontal line cuts at different $V_E$'s. The tilted shape indicates sizable cross-talk between $V_E$ and $V_P$, possibly due to the asymmetric gate spacing's. The labeled numbers are plateau values in units of $2e^2/h$. The blue arrow points to the $E_{1,2}$ degeneracy, where the "1" plateau is absent (see the blue curve in the lower panel). The black and green arrows point to the $E_{2,3}$ degeneracy, where the "2" plateau is absent (see the black and green curves in the lower panel). The red line cut in the lower panel is the case that the degeneracy between $E_1$, $E_2$, and $E_3$ are lifted, as both the "1" and "2" plateaus can be resolved. The "3" plateau is, however, absent, suggesting the $E_{3,4}$ degeneracy, see the red arrow in the upper panel. (d) $G$ vs $V_{top}$ and $V_{bottom}$ at $B = 0$ T. $V = 0$ mV. Lower panel, four line cuts. For larger $V_{top}$'s, the "2" plateau is absent (see the black and red line cuts), suggesting $E_{1,3}$ degeneracy. In the blue line cut, the "2" plateau emerges but the "1" plateau disappears, suggesting the $E_{1,2}$ degeneracy. For the green line cut, both "1" and "2" plateaus are present. (e) $G$ vs $V$ and $V_{bottom}$ at $B = 0$ T. $V_{top} = 3.8$ V. Lower panel, zero-bias line cut. The absence of the "2" plateau/diamond indicates the $E_{2,3}$ degeneracy. (f) $V_{top}$ is changed to -0.4 V, and the $E_{2,3}$ degeneracy is lifted, evidenced by the appearance of the "2" plateau/diamond. (g-i) $B$ scans along three axes. $V = 0$ mV. $V_{top} = 3.8$ V. The green arrow points to the degenerate case where the conductance plateau goes from "0.5" directly to "2". This change of $1.5 \times 2e^2/h$ suggests the presence of both types of degeneracy ($B$-induced and geometry symmetry-induced): $E_{1\downarrow} = E_{2\uparrow,3\uparrow}$. Contact resistance $R_{contact} = 650$ $\Omega$.



\end{document}